\documentclass[reprint,english,aip,bibnotes]{revtex4-1}
\usepackage{graphicx}
\usepackage[T1]{fontenc}
\usepackage[latin9]{inputenc}
\usepackage{amstext}
\usepackage{amssymb}
\usepackage{multirow}
\usepackage{dcolumn}
\usepackage{color}
\usepackage{amsmath}
\usepackage{babel}
\makeatletter

\begin{document}

\title{Design and characterization of a lumped element single-ended superconducting
microwave parametric amplifier with on-chip flux bias line}

\author{J.Y. Mutus}\thanks{These authors contributed equally to this work}
\affiliation{Department of Physics, University of California, Santa Barbara, California
93106-9530, USA}
\author{T.C. White}\thanks{These authors contributed equally to this work}
\affiliation{Department of Physics, University of California, Santa Barbara, California 93106-9530, USA}
\author{E. Jeffrey}
\affiliation{Department of Physics, University of California, Santa Barbara, California 93106-9530, USA}
\author{D. Sank}
\affiliation{Department of Physics, University of California, Santa Barbara, California 93106-9530, USA}
\author{R. Barends}
\affiliation{Department of Physics, University of California, Santa Barbara, California 93106-9530, USA}
\author{J. Bochmann}
\affiliation{Department of Physics, University of California, Santa Barbara, California 93106-9530, USA}
\author{Yu Chen}
\affiliation{Department of Physics, University of California, Santa Barbara, California 93106-9530, USA}
\author{Z. Chen}
\affiliation{Department of Physics, University of California, Santa Barbara, California 93106-9530, USA}
\author{B. Chiaro}
\affiliation{Department of Physics, University of California, Santa Barbara, California 93106-9530, USA}
\author{A. Dunsworth}
\affiliation{Department of Physics, University of California, Santa Barbara, California 93106-9530, USA}
\author{J. Kelly}
\affiliation{Department of Physics, University of California, Santa Barbara, California 93106-9530, USA}
\author{A. Megrant}
\affiliation{Department of Physics, University of California, Santa Barbara, California 93106-9530, USA}
\affiliation{Department of Materials, University of California, Santa Barbara, California 93106, USA}
\author{C. Neill}
\affiliation{Department of Physics, University of California, Santa Barbara, California 93106-9530, USA}
\author{P.J.J. O'Malley}
\affiliation{Department of Physics, University of California, Santa Barbara, California 93106-9530, USA}
\author{P. Roushan}
\affiliation{Department of Physics, University of California, Santa Barbara, California 93106-9530, USA}
\author{A. Vainsencher}
\affiliation{Department of Physics, University of California, Santa Barbara, California 93106-9530, USA}
\author{J. Wenner}
\affiliation{Department of Physics, University of California, Santa Barbara, California 93106-9530, USA}

\author{I. Siddiqi}
\affiliation{Quantum Nanoelectronics Laboratory, Department of Physics, University of California,
Berkeley, California 94720, USA}

\author{R. Vijay}
\affiliation{Quantum Nanoelectronics Laboratory, Department of Physics, University of California,
Berkeley, California 94720, USA}
\affiliation{ Department of Condensed Matter Physics and Materials Science, Tata Institute of Fundamental Research, Mumbai 400005, INDIA}

\author{A.N. Cleland}
\affiliation{Department of Physics, University of California, Santa Barbara, California 93106-9530, USA}
\affiliation{California NanoSystems Institute, University of California, Santa Barbara, CA 93106-9530, USA}
\author{John M. Martinis}
\email{martinis@physics.ucsb.edu}
\affiliation{Department of Physics, University of California, Santa Barbara, California 93106-9530, USA}
\affiliation{California NanoSystems Institute, University of California, Santa Barbara, CA 93106-9530, USA}

\date{\today}
\begin{abstract}
We demonstrate a lumped-element Josephson parametric amplifier, using a single-ended design that includes an on-chip, high-bandwidth flux bias line. The amplifier can be pumped into its region of parametric gain through either the input port or through the flux bias line. Broadband amplification is achieved at a tunable frequency $\omega/2 \pi$ between 5 to 7 GHz with quantum-limited noise performance, a gain-bandwidth product greater than 500 MHz, and an input saturation power in excess of -120 dBm. The bias line allows fast frequency tuning of the amplifier, with variations of hundreds of MHz over time scales shorter than 10 ns.
\end{abstract}

\maketitle


Low-power dispersive measurement of superconducting
microwave resonators has become an important tool for
applications ranging from the search for dark matter\cite{axion},
quantum-limited measurements of mechanical resonators\cite{regal:optomechanics},
and readout of superconducting qubits, where single-shot
sensitivity is desirable\cite{wallraff:dispersive,vijay:qJumps}. These measurements are typically performed using commercial cryogenic
high-electron-mobility transistor (HEMT) amplifiers\cite{bradley:HEMT},
which have several GHz of instantaneous bandwidth but add many
photons of noise to the measurement signal. Recently a number
of pre-amplifiers have been developed that achieve high gain with
near-quantum-limited performance, including DC SQUID amplifiers\cite{spietz:squidAmp},
the superconducting low-inductance undulating galvanometer (SLUG) amplifier\cite{hover:slug}, TiN traveling wave parametric
amplifiers\cite{eom:tinparamp}, and Josephson junction-based parametric
amplifiers\cite{yurke:originalJPA,siddiqi:JBA,castellanos:JPA,hatridge:LJPA,abdo:jpc,roch:JPC}.

In this work we present a lumped-element Josephson
parametric amplifier (LJPA), with a single-ended design and a
high-bandwidth on-chip bias line, based on the approach in Ref.
12. The simple single-ended design, fabricated using a multi-layer fabrication process, eliminates the need for a hybrid coupler used in differential designs. The amplifier can be pumped either through its input port or through the bias line, using a number of different operating
modes, making the device easy to adapt to a variety of
applications. We find this design yields wide bandwidth,
relatively high saturation power, and excellent noise
performance.


\begin{figure}
\includegraphics[height = 6cm]{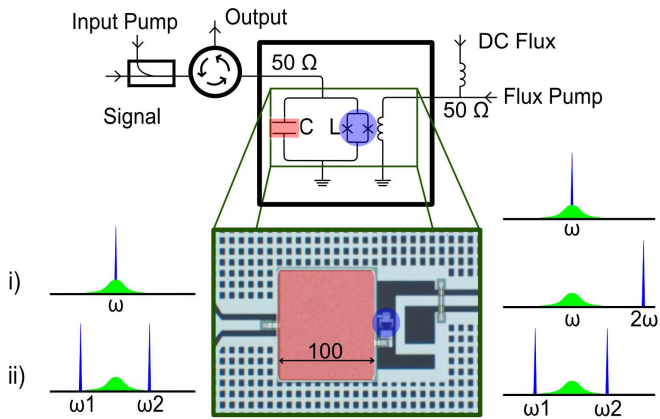} 
\caption{(Color online) Design of the paramp, where the signal is amplified through reflection off the non-linear resonant circuit. A circulator is used to separate incoming signals from amplified outgoing signals.  Top details the layout of the Josephson LC resonant circuit with input port (left) and bias line (right). Bottom shows a micrograph of the device; false color indicates the parallel plate capacitor (red square) and the SQUID (blue circle), adjacent to the bias line. A DC current, applied via the bias line, changes the coupled flux and tunes the resonant frequency of the amplifier. The device can be pumped through either the input port via the directional coupler or through the flux-pump port. The pump (blue, dark) and response (green, light) graphs display the five possible pump modes coupled to their respective terminals: (i) $\omega$-input, (ii) sideband-input, (iii) $\omega$-flux, (iv) 2$\omega$-flux, and (v) sideband-flux pumping.}
\label{figure:circuit} 
\end{figure}

Junction based superconducting paramps, regardless of design, depend on frequency mixing via the nonlinear Josephson inductance.  A sufficiently large pump tone can drive the circuit into a non-linear regime where energy couples from the pump to other tones within the device bandwidth.  Parametric amplification occurs when the pump ($\omega_{p}$), at the correct frequency and amplitude, transfers energy to a signal ($\omega_{s}$) and idler ($\omega_{i}$) tone.  Depending on the design, a paramp can operate as either a three-wave mixing amplifier, where ($\omega_{p} = \omega_{s} + \omega_{i}$) with typically $\omega_{p} \approx 2\omega_{s}$, or four-wave mixing amplifier, where ($2\omega_{p} = \omega_{s} + \omega_{i}$) with typically $\omega_{p} \approx \omega_{s}$.  In the degenerate output case ($\omega_{s} = \omega_{i}$) the signal and idler responses interfere to amplify only one quadrature, making phase sensitive operation possible with no added noise\cite{caves:qNoise}. In the more general non-degenerate output case ($\omega_{s} \neq \omega_{i}$), the phase of the amplified signal is preserved but an additional half photon of quantum noise is mixed into the signal response from the detuned idler frequency. In either form of amplification the circuit remains superconducting and dissipates minimal energy internally. Thus the half photon (degenerate) or whole photon (non-degenerate) of noise from quantum fluctuations is the dominant source of system noise\cite{caves:qNoise}.

Existing paramps, using both three-\cite{roch:JPC} and four-wave mixing\cite{siddiqi:JBA,castellanos:JPA,hatridge:LJPA}, operate in reflection mode: an incoming signal reflects off the amplifier, producing the outgoing amplified signal and idler tones. A microwave circulator separates the incoming from the outgoing signals, and provides a 50 $\Omega$ matched environment to eliminate standing waves at the input.  

The pump mode also dictates the hardware used in the signal path of the amplifier. For four-wave mixing ($\omega_{p} \approx \omega_{s}$), the large amplitude pump tone is combined with the signal using a directional coupler.  In this case additional isolating hardware (typically a circulator) is needed to prevent the reflected pump tone from perturbing the signal source (e.g. a qubit) in the measurement line. This hardware can cause loss of signal and decreased system quantum efficiency. In principle this can be avoided by using two equally detuned pump tones\cite{kamal:dp} (sideband pumping in Fig. \ref{figure:circuit} (ii, v)), but both tones must be precisely balanced to optimize paramp performance, making tuning the amplifier bias more complicated.  Three-wave mixing ($\omega_{p} \approx 2\omega_{s}$), where a single pump tone (Fig. \ref{figure:circuit} (iv)) is applied to the flux port, naturally avoids these constraints. Use of a single tone simplifies operation of the paramp, while pumping at twice the signal frequency eliminates the need for a directional coupler and naturally separates the pump and measurement signals.  
 
\begin{figure*}
\includegraphics[width = \textwidth]{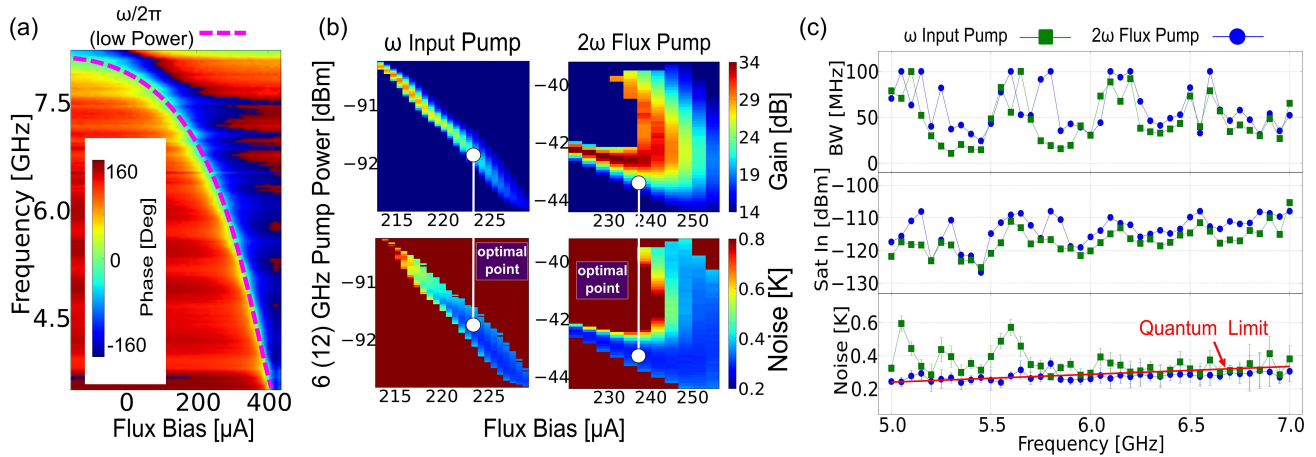}
\caption{(Color online) (a) Reflected phase of a low-power microwave signal vs. signal frequency ($\omega/2\pi$) over the range of a half flux quantum of DC flux bias. The dashed line (zero phase) corresponds to the linear (low power) resonant frequency and demonstrates a tunable range from 4 to 8 GHz.  (b) Power gain and system noise temperature (both referenced to input of directional coupler in Fig. \ref{figure:circuit}) vs. DC flux bias and pump power, for both $\omega$-input and 2$\omega$-flux pump modes. With 2$\omega$-flux pumping the device operates over a larger DC flux range and exhibits an additional branch where the amplifier operates with higher bandwidth at the cost of added noise.  For both modes, the optimal point (white circle) was chosen to maximize bandwidth and saturation power while maintaining large enough gain (22 dB) to ensure quantum-limited noise performance.  (c) Bandwidth,  input saturation power and system noise temperature vs. frequency,  for both pump modes.  Data was taken after tuning up the amplifier to its optimal point (as shown in (b)) every 50 MHz from 5 to 7 GHz. We see structure in the frequency dependence of the bandwidth and saturation power, which provides evidence for impedance variations in the microwave environment. While the noise temperature of the amplifier in the phase-preserving mode, obtained using both pump modes, scales with the frequency-dependent quantum noise (red line) given by $\hbar \omega /k_{B}$, 2$\omega$-flux pumping exhibits far less deviation from the quantum limit than $\omega$-input pumping.}
\label{figure:data} 
\end{figure*} 
 
In our single-ended device, as shown in the schematic and optical micrograph in Fig. \ref{figure:circuit}, the paramp resonant circuit consists of a SQUID loop with flux tunable inductance shunted by a parallel plate capacitor, with a resonant frequency in the 4-8 GHz range. The resonator is directly connected to the signal and ground of the 50 $\Omega$ input.  We tune the resonant frequency of the circuit by applying flux to the SQUID loop through an on-chip bias line, which is designed as a 50 $\Omega$ coplanar waveguide (CPW). 

This design leverages existing multilayer fabrication techniques first developed for use in the Josephson phase qubit\cite{martinis:dielectric}.  Low-loss amorphous silicon dielectric and low-impedance vias allow for 3-D routing of signal wires. We use these features to make a compact parallel plate capacitor and crossovers that eliminate CPW slot line modes.  These devices have a capacitance $C = 4.2$ pF, a stray inductance $L_{s} = 24$ pH, and an unbiased junction inductance $L_{j} = 68$ pH yielding a resonant frequency $(2\pi \sqrt{C(L_{s}+L_{j})})^{-1} = 8.1$ GHz with a coupled quality factor ($Q$) of 10.  The resonant frequency can be tuned from 4 to 8 GHz, shown in Fig. \ref{figure:data}(a), with a DC flux-bias current applied through the bias line with a mutual inductance $M = 1.4$ pH.  These devices are mass produced using wafer-scale fabrication, which yields hundreds of devices in parallel; we observe nominally identical performance between paramps from different chips.

This paramp, depending on which pump port is used and on the DC flux bias, can be operated as either a three-wave or four-wave mixing amplifier.  It can be operated as a four-wave mixing amplifier by the application of either a single or two detuned pump tones at the input port, which we refer to as $\omega$-input pumping (\ref{figure:circuit} (i)) and sideband-input pumping (\ref{figure:circuit} (ii)) respectively.  In addition, the paramp can be operated by driving RF flux through the SQUID loop with the high-bandwidth flux bias line.  This allows us to operate the device as either a four-wave amplifier with $\omega$-flux pumping (\ref{figure:circuit} (iii)) or a three-wave amplifier with 2$\omega$-flux pumping (\ref{figure:circuit} (iv)).  It is also possible to operate the device with sideband-flux pumping (\ref{figure:circuit} (v)), however more effort is required to tune the power of the pump tones on the flux-bias line. Figure \ref{figure:circuit} enumerates the five pump modes and indicates how each pump enters the circuit and the qualitative relationship between pump and amplified signal.  The ability to use all five modes allows the amplifier to be tailored to the requirements of a specific experiment.


LJPA-style paramps, using $\omega$-input pumping, have achieved gains greater than 25 dB and a gain-bandwidth product in the hundreds of MHz\cite{hatridge:LJPA}. This gain-bandwidth product has to date proven to be an order of magnitude higher than paramps based on other resonant circuits\cite{castellanos:JPAPerformance,yamamoto:fluxDrive}.  In our device we are able to replicate or exceed this performance with 2$\omega$-flux pumping. With our device we can identify the effect of pump mode on performance, using the same device under identical conditions. Among the five possible modes, we thoroughly investigated 2$\omega$-flux and $\omega$-input pumping, because these modes are the easiest to use and yield the best performance.

We compared the $\omega$-input and 2$\omega$-flux pump modes by measuring amplifier performance vs. frequency, pump power and detuning, shown in Fig. \ref{figure:data} (b-c). For a given LJPA-style amplifier, performance depends on pump power and the detuning between the pump frequency and the low-power resonant frequency\cite{siddiqi:JBAProperties}.  We changed the detuning by varying the DC flux bias of the amplifier while keeping the pump frequency constant.  Figure \ref{figure:data}(a) demonstrates the relationship between DC flux bias and the resonant frequency of the paramp, where we can tune from 4 to 8 GHz by applying between -300 and 400 $\mu$A of bias current, corresponding to a half-quantum of coupled flux.  We characterized device performance using gain, bandwidth, saturation power, and system noise temperature. At each frequency, power, and detuning, we measured the transmission and noise power vs. frequency with the pump off, then we re-measured the same quantities, as well as transmission power vs. signal power, with the pump on. The gain was calculated as the increase in transmission power near the pump frequency, the bandwidth as the full width at half maximum (FWHM) of the gain as a function of frequency, and the saturation power as the 1 dB compression point in gain vs. signal power.  These quantities were only measured for gains larger than 14 dB, below which the paramp does not overcome the noise added by the HEMT.  Here bandwidth refers to the full available bandwidth (for a given power and detuning) for constant wave (CW) signals.  When measuring pulsed (wide-band) signals, only half of the FWHM bandwidth can be used without generating distortion from mixing between signal and idler tones.  Lastly, we calculated the noise temperature of the amplifier in the phase-preserving mode using the method of signal-to-noise ratio improvement\cite{hatridge:LJPA} over a standard HEMT amplifier, at frequencies slightly detuned from the amplifier center frequency. Using a Y-factor measurement\cite{pozar} with a heated 50 $\Omega$ resistor installed on the mix plate of our dilution refrigerator (base temperature of 30 mK), the system noise temperature with only the HEMT was found to range from 1.8 to 2.6 K.

In general, we find that the same gain, bandwidth, and saturation can be achieved at multiple points in the pump power and detuning range, as shown explicitly for the gain in Fig. \ref{figure:data}(b).  We find 2$\omega$-flux pumping displays lower noise over a larger range of applied flux than does $\omega$-input pumping. Additionally,  2$\omega$-flux pumping features a branch at larger pump powers, where the amplifier operates with wider bandwidth, albeit at the cost of added noise.  When $\omega$-input pumping, the optimal noise performance is at the lowest power and detuning (bottom-right most point in Fig. \ref{figure:data}) for a given gain.  For 2$\omega$-flux pumping the entire lower branch exhibits quantum-limited performance.  We ensured operation on the low noise branch, by tuning to gains larger than 30 dB, which do not exist on the other branch.  Then for consistency, we chose the lowest power and detuning on the low noise branch which achieves the desired gain.

Previous studies have shown that LJPA performance is strongly dependent on the impedance of the environment in which the paramp circuit is embedded\cite{manucharyan:microEmbedding}.  As this impedance varies with frequency, we characterized this effect by measuring amplifier performance as a function of frequency.  Using the data from Fig. \ref{figure:data}(b) we implemented a software algorithm that tunes the amplifier to an optimal point in the parameter space, chosen as the smallest pump power and detuning that achieves 22 dB of gain with near quantum-limited noise. In this way we could ensure consistency between frequency measurements, eliminate experimenter bias, and automate the procedure.  The gain, bandwidth, saturation power and noise were measured using this technique from 5 to 7 GHz, shown in Fig. \ref{figure:data} (c). 

The data in Fig. \ref{figure:data}(c) display oscillations in the paramp bandwidth ranging from 30 MHz to 100 MHz, with an average of about 50 MHz.  The saturation power, which scales with pump power, also exhibits  similar oscillations of several dB, with the average increasing steadily from -125 dBm at 5 GHz to -110 dBm at 7 GHz.  The average saturation power scales with frequency because the pump power and frequency both scale with SQUID critical current.  The oscillations are evidence of variations in the impedance of the microwave environment, because both bandwidth and pump power are strongly dependent on the coupling $Q$ of the circuit\cite{manucharyan:microEmbedding}.  We observe a larger than average input saturation power in this device resulting from lower than average $Q$ as well as stray geometric inductance.  Stray inductance weakens the nonlinear response of the circuit requiring more pump power to operate\cite{levenson:nonlinear}.  For both pump modes, the noise temperature of the amplifier scales with the frequency-dependent quantum noise, but 2$\omega$-flux pumping deviates far less from the quantum limit than $\omega$-input pumping, especially at lower frequencies.   


\begin{figure}
\includegraphics[height = 6cm]{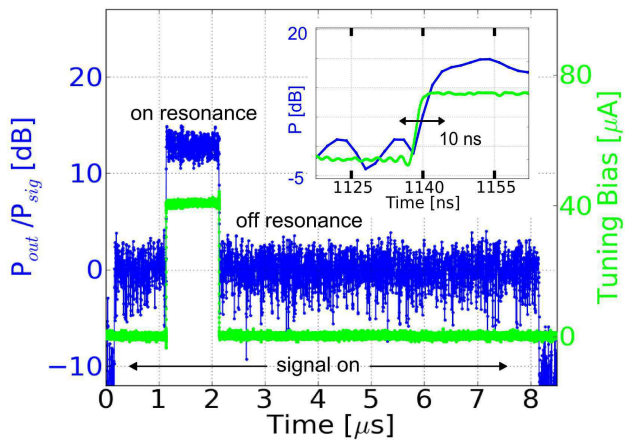} 
\caption{(Color online) Demonstration of dynamic frequency tuning of the amplifier. The output power normalized to the signal power (plotted in blue, dark) corresponds to the paramp gain, measured by mixing to DC and filtering.  Starting at 0.2 $\mu$s a small amplitude 5.98 GHz tone is applied to the paramp for 8 $\mu$s.   Using a 1 Giga-sample per second (Gsps) arbitrary waveform generator (AWG), the bias (plotted in green, light) is pulsed to 40 $\mu$A for 1 $\mu$s bringing the paramp on resonance and amplifying the 5.98 GHz tone.  Insert shows the leading edge of power gain jump on a 60 ns time scale.}
\label{figure:pulse} 
\end{figure}

In addition to allowing for multiple pump modes, the high bandwidth of the bias line can be used to rapidly tune the resonant frequency of the amplifier by varying the DC flux bias.  This can enable one paramp to sequentially monitor multiple signals widely spaced in frequency, e.g. for multiplexed resonator or qubit readout\cite{day:broadband, yates:fastmultiplex, chen:multiplexed, barends:xmon, groen:partial}. The paramp should respond to a change in its resonant frequency over a time scale limited only by the paramp's quality factor.  We verified this in 2$\omega$-flux pump mode by tuning the paramp to yield a gain of 14 dB at 5.98 GHz for a $2\omega / 2\pi = 12 $ GHz pump tone, corresponding to a paramp center frequency of about 6 GHz.  We then decreased the bias current by 40 $\mu$A, corresponding to a paramp frequency of about 6.4 GHz, and effectively reducing its gain at 6 GHz to unity.  To test the paramp time-domain response we then applied a 1 $\mu$s duration, 40 $\mu$A pulse of current generated by a 1 Giga-sample per second (Gsps) arbitrary wave form generator (AWG), and monitored the time-dependent gain of a 5.98 GHz signal tone, as shown in Fig. \ref{figure:pulse}.

The normalized paramp output power is shown in blue (dark) along with the AWG pulsed bias current in green (light). The time-domain signal was digitally mixed to DC and the idler tone and extra noise filtered. With the paramp off-resonant, the signal tone is only amplified by the HEMT and following amplifiers. During the current pulse, we see the gain increase suddenly by about 14 dB, in tandem with the DC current.  The inset shows a fine time-scale plot of the rising pulse edge.  These data show that the resonant frequency can be changed by several hundred MHz in less than 10 ns.


In summary, we have demonstrated a single-ended LJPA with gain exceeding 22 dB, gain-bandwidth product greater than 500 MHz, a saturation power greater than -120 dBm, and near quantum limited performance over an operating frequency from 5 to 7 GHz. This versatile device also allows for the comparison of three- and four-wave mixing using $\omega$-input and 2$\omega$-flux pumping modes. While the performance of the amplifier is similar for both pump modes, 2$\omega$-flux pumping offers better noise performance.  Since 2$\omega$-flux pumping also requires fewer components in the signal path, it is the preferred mode of operation. Lastly, we have demonstrated the inclusion of an on-chip bias line allows us to rapidly vary the resonant frequency of the amplifier.  This could be used to read-out widely-spaced (in frequency) signals using successive measurements separated by only $\sim$10 ns.

This work was supported by the Office of the Director of National Intel-ligence (ODNI), Intelligence Advanced Research Projects Activity (IARPA), through the Army Research Office grant W911NF-10-1-0334 and grant W911NF-11-1-0029. All statements of fact, opinion or conclusions contained herein are those of the authors and should not be con-strued as representing the official views or policies of IARPA, the ODNI, or the U.S. Government.  Devices were made at the UC Santa Barbara Nanofabrication Fa-cility, a part of the NSF-funded National Nanotechnol-ogy Infrastructure Network, and at the NanoStructures
Cleanroom Facility.


\end{document}